\documentstyle[pra,aps,psfig,epsfig]{revtex}
\begin{document}
\draft
\title{
\begin{flushright}
{\bf Preprint SSU-HEP-02/11\\
Samara State University}
\end{flushright}
\vspace{5mm}
\begin{center}
NUCLEAR STRUCTURE CORRECTIONS IN THE ENERGY\\
SPECTRA OF ELECTRONIC AND MUONIC DEUTERIUM\footnote{Talk presented
at the Conference "Physics of Fundamental Interactions" of the Nuclear
Physics Section of the Physics Department of RAS, ITEP, Moscow, 2-6 December,
2002}
\end{center}
}
\author{R.N.Faustov\footnote{E-mail:
faustov@theory.sinp.msu.ru}}
\address{117333, Moscow, Vavilov, 40, Scientific Council for Cybernetics RAS }
\author{A.P.Martynenko\footnote{E-mail:
mart@info.ssu.samara.ru}}
\address{443011, Samara, Pavlov, 1, Department of Theoretical Physics,
Samara State University}

\date{\today}

\maketitle

\begin{abstract}
The one-loop nuclear structure corrections of order $(Z\alpha)^5$
to the Lamb shift and hyperfine splitting of the deuterium are
calculated. The contribution of the deuteron structure effects
to the isotope shift (ep)-(ed), $(\mu p)-(\mu d)$ in the
interval $1S\div 2S$ is obtained on the basis of modern
experimental data on the deuteron electromagnetic form factors.
The comparison with the similar contributions to the Lamb shift
for the electronic and muonic hydrogen shows, that the relative
contribution due to the nucleus structure increases when passing
from the hydrogen to the deuterium.
\end{abstract}

\immediate\write16{<<WARNING: LINEDRAW macros work with emTeX-dvivers
                    and other drivers supporting emTeX \special's
                    (dviscr, dvihplj, dvidot, dvips, dviwin, etc.) >>}

\newdimen\Lengthunit       \Lengthunit  = 1.5cm
\newcount\Nhalfperiods     \Nhalfperiods= 9
\newcount\magnitude        \magnitude = 1000

\catcode`\*=11
\newdimen\L*   \newdimen\d*   \newdimen\d**
\newdimen\dm*  \newdimen\dd*  \newdimen\dt*
\newdimen\a*   \newdimen\b*   \newdimen\c*
\newdimen\a**  \newdimen\b**
\newdimen\xL*  \newdimen\yL*
\newdimen\rx*  \newdimen\ry*
\newdimen\tmp* \newdimen\linwid*

\newcount\k*   \newcount\l*   \newcount\m*
\newcount\k**  \newcount\l**  \newcount\m**
\newcount\n*   \newcount\dn*  \newcount\r*
\newcount\N*   \newcount\*one \newcount\*two  \*one=1 \*two=2
\newcount\*ths \*ths=1000
\newcount\angle*  \newcount\q*  \newcount\q**
\newcount\angle** \angle**=0
\newcount\sc*     \sc*=0

\newtoks\cos*  \cos*={1}
\newtoks\sin*  \sin*={0}

\catcode`\[=13

\def\rotate(#1){\advance\angle**#1\angle*=\angle**
\q**=\angle*\ifnum\q**<0\q**=-\q**\fi
\ifnum\q**>360\q*=\angle*\divide\q*360\multiply\q*360\advance\angle*-\q*\fi
\ifnum\angle*<0\advance\angle*360\fi\q**=\angle*\divide\q**90\q**=\q**
\def\sgcos*{+}\def\sgsin*{+}\relax
\ifcase\q**\or
 \def\sgcos*{-}\def\sgsin*{+}\or
 \def\sgcos*{-}\def\sgsin*{-}\or
 \def\sgcos*{+}\def\sgsin*{-}\else\fi
\q*=\q**
\multiply\q*90\advance\angle*-\q*
\ifnum\angle*>45\sc*=1\angle*=-\angle*\advance\angle*90\else\sc*=0\fi
\def[##1,##2]{\ifnum\sc*=0\relax
\edef\cs*{\sgcos*.##1}\edef\sn*{\sgsin*.##2}\ifcase\q**\or
 \edef\cs*{\sgcos*.##2}\edef\sn*{\sgsin*.##1}\or
 \edef\cs*{\sgcos*.##1}\edef\sn*{\sgsin*.##2}\or
 \edef\cs*{\sgcos*.##2}\edef\sn*{\sgsin*.##1}\else\fi\else
\edef\cs*{\sgcos*.##2}\edef\sn*{\sgsin*.##1}\ifcase\q**\or
 \edef\cs*{\sgcos*.##1}\edef\sn*{\sgsin*.##2}\or
 \edef\cs*{\sgcos*.##2}\edef\sn*{\sgsin*.##1}\or
 \edef\cs*{\sgcos*.##1}\edef\sn*{\sgsin*.##2}\else\fi\fi
\cos*={\cs*}\sin*={\sn*}\global\edef\gcos*{\cs*}\global\edef\gsin*{\sn*}}\relax
\ifcase\angle*[9999,0]\or
[999,017]\or[999,034]\or[998,052]\or[997,069]\or[996,087]\or
[994,104]\or[992,121]\or[990,139]\or[987,156]\or[984,173]\or
[981,190]\or[978,207]\or[974,224]\or[970,241]\or[965,258]\or
[961,275]\or[956,292]\or[951,309]\or[945,325]\or[939,342]\or
[933,358]\or[927,374]\or[920,390]\or[913,406]\or[906,422]\or
[898,438]\or[891,453]\or[882,469]\or[874,484]\or[866,499]\or
[857,515]\or[848,529]\or[838,544]\or[829,559]\or[819,573]\or
[809,587]\or[798,601]\or[788,615]\or[777,629]\or[766,642]\or
[754,656]\or[743,669]\or[731,681]\or[719,694]\or[707,707]\or
\else[9999,0]\fi}

\catcode`\[=12

\def\GRAPH(hsize=#1)#2{\hbox to #1\Lengthunit{#2\hss}}

\def\Linewidth#1{\global\linwid*=#1\relax
\global\divide\linwid*10\global\multiply\linwid*\mag
\global\divide\linwid*100\special{em:linewidth \the\linwid*}}

\Linewidth{.4pt}
\def\sm*{\special{em:moveto}}
\def\sl*{\special{em:lineto}}
\let\moveto=\sm*
\let\lineto=\sl*
\newbox\spm*   \newbox\spl*
\setbox\spm*\hbox{\sm*}
\setbox\spl*\hbox{\sl*}

\def\mov#1(#2,#3)#4{\rlap{\L*=#1\Lengthunit
\xL*=#2\L* \yL*=#3\L*
\xL*=\xscale\xL* \yL*=\yscale\yL*
\rx* \the\cos*\xL* \tmp* \the\sin*\yL* \advance\rx*-\tmp*
\ry* \the\cos*\yL* \tmp* \the\sin*\xL* \advance\ry*\tmp*
\kern\rx*\raise\ry*\hbox{#4}}}

\def\rmov*(#1,#2)#3{\rlap{\xL*=#1\yL*=#2\relax
\rx* \the\cos*\xL* \tmp* \the\sin*\yL* \advance\rx*-\tmp*
\ry* \the\cos*\yL* \tmp* \the\sin*\xL* \advance\ry*\tmp*
\kern\rx*\raise\ry*\hbox{#3}}}

\def\lin#1(#2,#3){\rlap{\sm*\mov#1(#2,#3){\sl*}}}

\def\arr*(#1,#2,#3){\rmov*(#1\dd*,#1\dt*){\sm*
\rmov*(#2\dd*,#2\dt*){\rmov*(#3\dt*,-#3\dd*){\sl*}}\sm*
\rmov*(#2\dd*,#2\dt*){\rmov*(-#3\dt*,#3\dd*){\sl*}}}}

\def\arrow#1(#2,#3){\rlap{\lin#1(#2,#3)\mov#1(#2,#3){\relax
\d**=-.012\Lengthunit\dd*=#2\d**\dt*=#3\d**
\arr*(1,10,4)\arr*(3,8,4)\arr*(4.8,4.2,3)}}}

\def\arrlin#1(#2,#3){\rlap{\L*=#1\Lengthunit\L*=.5\L*
\lin#1(#2,#3)\rmov*(#2\L*,#3\L*){\arrow.1(#2,#3)}}}

\def\dasharrow#1(#2,#3){\rlap{{\Lengthunit=0.9\Lengthunit
\dashlin#1(#2,#3)\mov#1(#2,#3){\sm*}}\mov#1(#2,#3){\sl*
\d**=-.012\Lengthunit\dd*=#2\d**\dt*=#3\d**
\arr*(1,10,4)\arr*(3,8,4)\arr*(4.8,4.2,3)}}}

\def\clap#1{\hbox to 0pt{\hss #1\hss}}

\def\ind(#1,#2)#3{\rlap{\L*=.1\Lengthunit
\xL*=#1\L* \yL*=#2\L*
\rx* \the\cos*\xL* \tmp* \the\sin*\yL* \advance\rx*-\tmp*
\ry* \the\cos*\yL* \tmp* \the\sin*\xL* \advance\ry*\tmp*
\kern\rx*\raise\ry*\hbox{\lower2pt\clap{$#3$}}}}

\def\sh*(#1,#2)#3{\rlap{\dm*=\the\n*\d**
\xL*=\xscale\dm* \yL*=\yscale\dm* \xL*=#1\xL* \yL*=#2\yL*
\rx* \the\cos*\xL* \tmp* \the\sin*\yL* \advance\rx*-\tmp*
\ry* \the\cos*\yL* \tmp* \the\sin*\xL* \advance\ry*\tmp*
\kern\rx*\raise\ry*\hbox{#3}}}

\def\calcnum*#1(#2,#3){\a*=1000sp\b*=1000sp\a*=#2\a*\b*=#3\b*
\ifdim\a*<0pt\a*-\a*\fi\ifdim\b*<0pt\b*-\b*\fi
\ifdim\a*>\b*\c*=.96\a*\advance\c*.4\b*
\else\c*=.96\b*\advance\c*.4\a*\fi
\k*\a*\multiply\k*\k*\l*\b*\multiply\l*\l*
\m*\k*\advance\m*\l*\n*\c*\r*\n*\multiply\n*\n*
\dn*\m*\advance\dn*-\n*\divide\dn*2\divide\dn*\r*
\advance\r*\dn*
\c*=\the\Nhalfperiods5sp\c*=#1\c*\ifdim\c*<0pt\c*-\c*\fi
\multiply\c*\r*\N*\c*\divide\N*10000}

\def\dashlin#1(#2,#3){\rlap{\calcnum*#1(#2,#3)\relax
\d**=#1\Lengthunit\ifdim\d**<0pt\d**-\d**\fi
\divide\N*2\multiply\N*2\advance\N*\*one
\divide\d**\N*\sm*\n*\*one\sh*(#2,#3){\sl*}\loop
\advance\n*\*one\sh*(#2,#3){\sm*}\advance\n*\*one
\sh*(#2,#3){\sl*}\ifnum\n*<\N*\repeat}}

\def\dashdotlin#1(#2,#3){\rlap{\calcnum*#1(#2,#3)\relax
\d**=#1\Lengthunit\ifdim\d**<0pt\d**-\d**\fi
\divide\N*2\multiply\N*2\advance\N*1\multiply\N*2\relax
\divide\d**\N*\sm*\n*\*two\sh*(#2,#3){\sl*}\loop
\advance\n*\*one\sh*(#2,#3){\kern-1.48pt\lower.5pt\hbox{\rm.}}\relax
\advance\n*\*one\sh*(#2,#3){\sm*}\advance\n*\*two
\sh*(#2,#3){\sl*}\ifnum\n*<\N*\repeat}}

\def\shl*(#1,#2)#3{\kern#1#3\lower#2#3\hbox{\unhcopy\spl*}}

\def\trianglin#1(#2,#3){\rlap{\toks0={#2}\toks1={#3}\calcnum*#1(#2,#3)\relax
\dd*=.57\Lengthunit\dd*=#1\dd*\divide\dd*\N*
\divide\dd*\*ths \multiply\dd*\magnitude
\d**=#1\Lengthunit\ifdim\d**<0pt\d**-\d**\fi
\multiply\N*2\divide\d**\N*\sm*\n*\*one\loop
\shl**{\dd*}\dd*-\dd*\advance\n*2\relax
\ifnum\n*<\N*\repeat\n*\N*\shl**{0pt}}}

\def\wavelin#1(#2,#3){\rlap{\toks0={#2}\toks1={#3}\calcnum*#1(#2,#3)\relax
\dd*=.23\Lengthunit\dd*=#1\dd*\divide\dd*\N*
\divide\dd*\*ths \multiply\dd*\magnitude
\d**=#1\Lengthunit\ifdim\d**<0pt\d**-\d**\fi
\multiply\N*4\divide\d**\N*\sm*\n*\*one\loop
\shl**{\dd*}\dt*=1.3\dd*\advance\n*\*one
\shl**{\dt*}\advance\n*\*one
\shl**{\dd*}\advance\n*\*two
\dd*-\dd*\ifnum\n*<\N*\repeat\n*\N*\shl**{0pt}}}

\def\w*lin(#1,#2){\rlap{\toks0={#1}\toks1={#2}\d**=\Lengthunit\dd*=-.12\d**
\divide\dd*\*ths \multiply\dd*\magnitude
\N*8\divide\d**\N*\sm*\n*\*one\loop
\shl**{\dd*}\dt*=1.3\dd*\advance\n*\*one
\shl**{\dt*}\advance\n*\*one
\shl**{\dd*}\advance\n*\*one
\shl**{0pt}\dd*-\dd*\advance\n*1\ifnum\n*<\N*\repeat}}

\def\l*arc(#1,#2)[#3][#4]{\rlap{\toks0={#1}\toks1={#2}\d**=\Lengthunit
\dd*=#3.037\d**\dd*=#4\dd*\dt*=#3.049\d**\dt*=#4\dt*\ifdim\d**>10mm\relax
\d**=.25\d**\n*\*one\shl**{-\dd*}\n*\*two\shl**{-\dt*}\n*3\relax
\shl**{-\dd*}\n*4\relax\shl**{0pt}\else
\ifdim\d**>5mm\d**=.5\d**\n*\*one\shl**{-\dt*}\n*\*two
\shl**{0pt}\else\n*\*one\shl**{0pt}\fi\fi}}

\def\d*arc(#1,#2)[#3][#4]{\rlap{\toks0={#1}\toks1={#2}\d**=\Lengthunit
\dd*=#3.037\d**\dd*=#4\dd*\d**=.25\d**\sm*\n*\*one\shl**{-\dd*}\relax
\n*3\relax\sh*(#1,#2){\xL*=\xscale\dd*\yL*=\yscale\dd*
\kern#2\xL*\lower#1\yL*\hbox{\sm*}}\n*4\relax\shl**{0pt}}}

\def\shl**#1{\c*=\the\n*\d**\d*=#1\relax
\a*=\the\toks0\c*\b*=\the\toks1\d*\advance\a*-\b*
\b*=\the\toks1\c*\d*=\the\toks0\d*\advance\b*\d*
\a*=\xscale\a*\b*=\yscale\b*
\rx* \the\cos*\a* \tmp* \the\sin*\b* \advance\rx*-\tmp*
\ry* \the\cos*\b* \tmp* \the\sin*\a* \advance\ry*\tmp*
\raise\ry*\rlap{\kern\rx*\unhcopy\spl*}}

\def\wlin*#1(#2,#3)[#4]{\rlap{\toks0={#2}\toks1={#3}\relax
\c*=#1\l*\c*\c*=.01\Lengthunit\m*\c*\divide\l*\m*
\c*=\the\Nhalfperiods5sp\multiply\c*\l*\N*\c*\divide\N*\*ths
\divide\N*2\multiply\N*2\advance\N*\*one
\dd*=.002\Lengthunit\dd*=#4\dd*\multiply\dd*\l*\divide\dd*\N*
\divide\dd*\*ths \multiply\dd*\magnitude
\d**=#1\multiply\N*4\divide\d**\N*\sm*\n*\*one\loop
\shl**{\dd*}\dt*=1.3\dd*\advance\n*\*one
\shl**{\dt*}\advance\n*\*one
\shl**{\dd*}\advance\n*\*two
\dd*-\dd*\ifnum\n*<\N*\repeat\n*\N*\shl**{0pt}}}

\def\wavebox#1{\setbox0\hbox{#1}\relax
\a*=\wd0\advance\a*14pt\b*=\ht0\advance\b*\dp0\advance\b*14pt\relax
\hbox{\kern9pt\relax
\rmov*(0pt,\ht0){\rmov*(-7pt,7pt){\wlin*\a*(1,0)[+]\wlin*\b*(0,-1)[-]}}\relax
\rmov*(\wd0,-\dp0){\rmov*(7pt,-7pt){\wlin*\a*(-1,0)[+]\wlin*\b*(0,1)[-]}}\relax
\box0\kern9pt}}

\def\rectangle#1(#2,#3){\relax
\lin#1(#2,0)\lin#1(0,#3)\mov#1(0,#3){\lin#1(#2,0)}\mov#1(#2,0){\lin#1(0,#3)}}

\def\dashrectangle#1(#2,#3){\dashlin#1(#2,0)\dashlin#1(0,#3)\relax
\mov#1(0,#3){\dashlin#1(#2,0)}\mov#1(#2,0){\dashlin#1(0,#3)}}

\def\waverectangle#1(#2,#3){\L*=#1\Lengthunit\a*=#2\L*\b*=#3\L*
\ifdim\a*<0pt\a*-\a*\def\x*{-1}\else\def\x*{1}\fi
\ifdim\b*<0pt\b*-\b*\def\y*{-1}\else\def\y*{1}\fi
\wlin*\a*(\x*,0)[-]\wlin*\b*(0,\y*)[+]\relax
\mov#1(0,#3){\wlin*\a*(\x*,0)[+]}\mov#1(#2,0){\wlin*\b*(0,\y*)[-]}}

\def\calcparab*{\ifnum\n*>\m*\k*\N*\advance\k*-\n*\else\k*\n*\fi
\a*=\the\k* sp\a*=10\a*\b*\dm*\advance\b*-\a*\k*\b*
\a*=\the\*ths\b*\divide\a*\l*\multiply\a*\k*
\divide\a*\l*\k*\*ths\r*\a*\advance\k*-\r*\dt*=\the\k*\L*}

\def\arcto#1(#2,#3)[#4]{\rlap{\toks0={#2}\toks1={#3}\calcnum*#1(#2,#3)\relax
\dm*=135sp\dm*=#1\dm*\d**=#1\Lengthunit\ifdim\dm*<0pt\dm*-\dm*\fi
\multiply\dm*\r*\a*=.3\dm*\a*=#4\a*\ifdim\a*<0pt\a*-\a*\fi
\advance\dm*\a*\N*\dm*\divide\N*10000\relax
\divide\N*2\multiply\N*2\advance\N*\*one
\L*=-.25\d**\L*=#4\L*\divide\d**\N*\divide\L*\*ths
\m*\N*\divide\m*2\dm*=\the\m*5sp\l*\dm*\sm*\n*\*one\loop
\calcparab*\shl**{-\dt*}\advance\n*1\ifnum\n*<\N*\repeat}}

\def\arrarcto#1(#2,#3)[#4]{\L*=#1\Lengthunit\L*=.54\L*
\arcto#1(#2,#3)[#4]\rmov*(#2\L*,#3\L*){\d*=.457\L*\d*=#4\d*\d**-\d*
\rmov*(#3\d**,#2\d*){\arrow.02(#2,#3)}}}

\def\dasharcto#1(#2,#3)[#4]{\rlap{\toks0={#2}\toks1={#3}\relax
\calcnum*#1(#2,#3)\dm*=\the\N*5sp\a*=.3\dm*\a*=#4\a*\ifdim\a*<0pt\a*-\a*\fi
\advance\dm*\a*\N*\dm*
\divide\N*20\multiply\N*2\advance\N*1\d**=#1\Lengthunit
\L*=-.25\d**\L*=#4\L*\divide\d**\N*\divide\L*\*ths
\m*\N*\divide\m*2\dm*=\the\m*5sp\l*\dm*
\sm*\n*\*one\loop\calcparab*
\shl**{-\dt*}\advance\n*1\ifnum\n*>\N*\else\calcparab*
\sh*(#2,#3){\xL*=#3\dt* \yL*=#2\dt*
\rx* \the\cos*\xL* \tmp* \the\sin*\yL* \advance\rx*\tmp*
\ry* \the\cos*\yL* \tmp* \the\sin*\xL* \advance\ry*-\tmp*
\kern\rx*\lower\ry*\hbox{\sm*}}\fi
\advance\n*1\ifnum\n*<\N*\repeat}}

\def\*shl*#1{\c*=\the\n*\d**\advance\c*#1\a**\d*\dt*\advance\d*#1\b**
\a*=\the\toks0\c*\b*=\the\toks1\d*\advance\a*-\b*
\b*=\the\toks1\c*\d*=\the\toks0\d*\advance\b*\d*
\rx* \the\cos*\a* \tmp* \the\sin*\b* \advance\rx*-\tmp*
\ry* \the\cos*\b* \tmp* \the\sin*\a* \advance\ry*\tmp*
\raise\ry*\rlap{\kern\rx*\unhcopy\spl*}}

\def\calcnormal*#1{\b**=10000sp\a**\b**\k*\n*\advance\k*-\m*
\multiply\a**\k*\divide\a**\m*\a**=#1\a**\ifdim\a**<0pt\a**-\a**\fi
\ifdim\a**>\b**\d*=.96\a**\advance\d*.4\b**
\else\d*=.96\b**\advance\d*.4\a**\fi
\d*=.01\d*\r*\d*\divide\a**\r*\divide\b**\r*
\ifnum\k*<0\a**-\a**\fi\d*=#1\d*\ifdim\d*<0pt\b**-\b**\fi
\k*\a**\a**=\the\k*\dd*\k*\b**\b**=\the\k*\dd*}

\def\wavearcto#1(#2,#3)[#4]{\rlap{\toks0={#2}\toks1={#3}\relax
\calcnum*#1(#2,#3)\c*=\the\N*5sp\a*=.4\c*\a*=#4\a*\ifdim\a*<0pt\a*-\a*\fi
\advance\c*\a*\N*\c*\divide\N*20\multiply\N*2\advance\N*-1\multiply\N*4\relax
\d**=#1\Lengthunit\dd*=.012\d**
\divide\dd*\*ths \multiply\dd*\magnitude
\ifdim\d**<0pt\d**-\d**\fi\L*=.25\d**
\divide\d**\N*\divide\dd*\N*\L*=#4\L*\divide\L*\*ths
\m*\N*\divide\m*2\dm*=\the\m*0sp\l*\dm*
\sm*\n*\*one\loop\calcnormal*{#4}\calcparab*
\*shl*{1}\advance\n*\*one\calcparab*
\*shl*{1.3}\advance\n*\*one\calcparab*
\*shl*{1}\advance\n*2\dd*-\dd*\ifnum\n*<\N*\repeat\n*\N*\shl**{0pt}}}

\def\triangarcto#1(#2,#3)[#4]{\rlap{\toks0={#2}\toks1={#3}\relax
\calcnum*#1(#2,#3)\c*=\the\N*5sp\a*=.4\c*\a*=#4\a*\ifdim\a*<0pt\a*-\a*\fi
\advance\c*\a*\N*\c*\divide\N*20\multiply\N*2\advance\N*-1\multiply\N*2\relax
\d**=#1\Lengthunit\dd*=.012\d**
\divide\dd*\*ths \multiply\dd*\magnitude
\ifdim\d**<0pt\d**-\d**\fi\L*=.25\d**
\divide\d**\N*\divide\dd*\N*\L*=#4\L*\divide\L*\*ths
\m*\N*\divide\m*2\dm*=\the\m*0sp\l*\dm*
\sm*\n*\*one\loop\calcnormal*{#4}\calcparab*
\*shl*{1}\advance\n*2\dd*-\dd*\ifnum\n*<\N*\repeat\n*\N*\shl**{0pt}}}

\def\hr*#1{\L*=\xscale\Lengthunit\ifnum
\angle**=0\clap{\vrule width#1\L* height.1pt}\else
\L*=#1\L*\L*=.5\L*\rmov*(-\L*,0pt){\sm*}\rmov*(\L*,0pt){\sl*}\fi}

\def\shade#1[#2]{\rlap{\Lengthunit=#1\Lengthunit
\special{em:linewidth .001pt}\relax
\mov(0,#2.05){\hr*{.994}}\mov(0,#2.1){\hr*{.980}}\relax
\mov(0,#2.15){\hr*{.953}}\mov(0,#2.2){\hr*{.916}}\relax
\mov(0,#2.25){\hr*{.867}}\mov(0,#2.3){\hr*{.798}}\relax
\mov(0,#2.35){\hr*{.715}}\mov(0,#2.4){\hr*{.603}}\relax
\mov(0,#2.45){\hr*{.435}}\special{em:linewidth \the\linwid*}}}

\def\dshade#1[#2]{\rlap{\special{em:linewidth .001pt}\relax
\Lengthunit=#1\Lengthunit\if#2-\def\t*{+}\else\def\t*{-}\fi
\mov(0,\t*.025){\relax
\mov(0,#2.05){\hr*{.995}}\mov(0,#2.1){\hr*{.988}}\relax
\mov(0,#2.15){\hr*{.969}}\mov(0,#2.2){\hr*{.937}}\relax
\mov(0,#2.25){\hr*{.893}}\mov(0,#2.3){\hr*{.836}}\relax
\mov(0,#2.35){\hr*{.760}}\mov(0,#2.4){\hr*{.662}}\relax
\mov(0,#2.45){\hr*{.531}}\mov(0,#2.5){\hr*{.320}}\relax
\special{em:linewidth \the\linwid*}}}}

\def\vdot{\rlap{\kern-1.9pt\lower1.8pt\hbox{$\scriptstyle\bullet$}}}
\def\vtimes{\rlap{\kern-3pt\lower1.8pt\hbox{$\scriptstyle\times$}}}
\def\vDot{\rlap{\kern-2.3pt\lower2.7pt\hbox{$\bullet$}}}
\def\vTimes{\rlap{\kern-3.6pt\lower2.4pt\hbox{$\times$}}}

\def\arc(#1)[#2,#3]{{\k*=#2\l*=#3\m*=\l*
\advance\m*-6\ifnum\k*>\l*\relax\else
{\rotate(#2)\mov(#1,0){\sm*}}\loop
\ifnum\k*<\m*\advance\k*5{\rotate(\k*)\mov(#1,0){\sl*}}\repeat
{\rotate(#3)\mov(#1,0){\sl*}}\fi}}

\def\dasharc(#1)[#2,#3]{{\k**=#2\n*=#3\advance\n*-1\advance\n*-\k**
\L*=1000sp\L*#1\L* \multiply\L*\n* \multiply\L*\Nhalfperiods
\divide\L*57\N*\L* \divide\N*2000\ifnum\N*=0\N*1\fi
\r*\n*  \divide\r*\N* \ifnum\r*<2\r*2\fi
\m**\r* \divide\m**2 \l**\r* \advance\l**-\m** \N*\n* \divide\N*\r*
\k**\r* \multiply\k**\N* \dn*\n* \advance\dn*-\k** \divide\dn*2\advance\dn*\*one
\r*\l** \divide\r*2\advance\dn*\r* \advance\N*-2\k**#2\relax
\ifnum\l**<6{\rotate(#2)\mov(#1,0){\sm*}}\advance\k**\dn*
{\rotate(\k**)\mov(#1,0){\sl*}}\advance\k**\m**
{\rotate(\k**)\mov(#1,0){\sm*}}\loop
\advance\k**\l**{\rotate(\k**)\mov(#1,0){\sl*}}\advance\k**\m**
{\rotate(\k**)\mov(#1,0){\sm*}}\advance\N*-1\ifnum\N*>0\repeat
{\rotate(#3)\mov(#1,0){\sl*}}\else\advance\k**\dn*
\arc(#1)[#2,\k**]\loop\advance\k**\m** \r*\k**
\advance\k**\l** {\arc(#1)[\r*,\k**]}\relax
\advance\N*-1\ifnum\N*>0\repeat
\advance\k**\m**\arc(#1)[\k**,#3]\fi}}

\def\triangarc#1(#2)[#3,#4]{{\k**=#3\n*=#4\advance\n*-\k**
\L*=1000sp\L*#2\L* \multiply\L*\n* \multiply\L*\Nhalfperiods
\divide\L*57\N*\L* \divide\N*1000\ifnum\N*=0\N*1\fi
\d**=#2\Lengthunit \d*\d** \divide\d*57\multiply\d*\n*
\r*\n*  \divide\r*\N* \ifnum\r*<2\r*2\fi
\m**\r* \divide\m**2 \l**\r* \advance\l**-\m** \N*\n* \divide\N*\r*
\dt*\d* \divide\dt*\N* \dt*.5\dt* \dt*#1\dt*
\divide\dt*1000\multiply\dt*\magnitude
\k**\r* \multiply\k**\N* \dn*\n* \advance\dn*-\k** \divide\dn*2\relax
\r*\l** \divide\r*2\advance\dn*\r* \advance\N*-1\k**#3\relax
{\rotate(#3)\mov(#2,0){\sm*}}\advance\k**\dn*
{\rotate(\k**)\mov(#2,0){\sl*}}\advance\k**-\m**\advance\l**\m**\loop\dt*-\dt*
\d*\d** \advance\d*\dt*
\advance\k**\l**{\rotate(\k**)\rmov*(\d*,0pt){\sl*}}%
\advance\N*-1\ifnum\N*>0\repeat\advance\k**\m**
{\rotate(\k**)\mov(#2,0){\sl*}}{\rotate(#4)\mov(#2,0){\sl*}}}}

\def\wavearc#1(#2)[#3,#4]{{\k**=#3\n*=#4\advance\n*-\k**
\L*=4000sp\L*#2\L* \multiply\L*\n* \multiply\L*\Nhalfperiods
\divide\L*57\N*\L* \divide\N*1000\ifnum\N*=0\N*1\fi
\d**=#2\Lengthunit \d*\d** \divide\d*57\multiply\d*\n*
\r*\n*  \divide\r*\N* \ifnum\r*=0\r*1\fi
\m**\r* \divide\m**2 \l**\r* \advance\l**-\m** \N*\n* \divide\N*\r*
\dt*\d* \divide\dt*\N* \dt*.7\dt* \dt*#1\dt*
\divide\dt*1000\multiply\dt*\magnitude
\k**\r* \multiply\k**\N* \dn*\n* \advance\dn*-\k** \divide\dn*2\relax
\divide\N*4\advance\N*-1\k**#3\relax
{\rotate(#3)\mov(#2,0){\sm*}}\advance\k**\dn*
{\rotate(\k**)\mov(#2,0){\sl*}}\advance\k**-\m**\advance\l**\m**\loop\dt*-\dt*
\d*\d** \advance\d*\dt* \dd*\d** \advance\dd*1.3\dt*
\advance\k**\r*{\rotate(\k**)\rmov*(\d*,0pt){\sl*}}\relax
\advance\k**\r*{\rotate(\k**)\rmov*(\dd*,0pt){\sl*}}\relax
\advance\k**\r*{\rotate(\k**)\rmov*(\d*,0pt){\sl*}}\relax
\advance\k**\r*
\advance\N*-1\ifnum\N*>0\repeat\advance\k**\m**
{\rotate(\k**)\mov(#2,0){\sl*}}{\rotate(#4)\mov(#2,0){\sl*}}}}

\def\gmov*#1(#2,#3)#4{\rlap{\L*=#1\Lengthunit
\xL*=#2\L* \yL*=#3\L*
\rx* \gcos*\xL* \tmp* \gsin*\yL* \advance\rx*-\tmp*
\ry* \gcos*\yL* \tmp* \gsin*\xL* \advance\ry*\tmp*
\rx*=\xscale\rx* \ry*=\yscale\ry*
\xL* \the\cos*\rx* \tmp* \the\sin*\ry* \advance\xL*-\tmp*
\yL* \the\cos*\ry* \tmp* \the\sin*\rx* \advance\yL*\tmp*
\kern\xL*\raise\yL*\hbox{#4}}}

\def\rgmov*(#1,#2)#3{\rlap{\xL*#1\yL*#2\relax
\rx* \gcos*\xL* \tmp* \gsin*\yL* \advance\rx*-\tmp*
\ry* \gcos*\yL* \tmp* \gsin*\xL* \advance\ry*\tmp*
\rx*=\xscale\rx* \ry*=\yscale\ry*
\xL* \the\cos*\rx* \tmp* \the\sin*\ry* \advance\xL*-\tmp*
\yL* \the\cos*\ry* \tmp* \the\sin*\rx* \advance\yL*\tmp*
\kern\xL*\raise\yL*\hbox{#3}}}

\def\Earc(#1)[#2,#3][#4,#5]{{\k*=#2\l*=#3\m*=\l*
\advance\m*-6\ifnum\k*>\l*\relax\else\def\xscale{#4}\def\yscale{#5}\relax
{\angle**0\rotate(#2)}\gmov*(#1,0){\sm*}\loop
\ifnum\k*<\m*\advance\k*5\relax
{\angle**0\rotate(\k*)}\gmov*(#1,0){\sl*}\repeat
{\angle**0\rotate(#3)}\gmov*(#1,0){\sl*}\relax
\def\xscale{1}\def\yscale{1}\fi}}

\def\dashEarc(#1)[#2,#3][#4,#5]{{\k**=#2\n*=#3\advance\n*-1\advance\n*-\k**
\L*=1000sp\L*#1\L* \multiply\L*\n* \multiply\L*\Nhalfperiods
\divide\L*57\N*\L* \divide\N*2000\ifnum\N*=0\N*1\fi
\r*\n*  \divide\r*\N* \ifnum\r*<2\r*2\fi
\m**\r* \divide\m**2 \l**\r* \advance\l**-\m** \N*\n* \divide\N*\r*
\k**\r*\multiply\k**\N* \dn*\n* \advance\dn*-\k** \divide\dn*2\advance\dn*\*one
\r*\l** \divide\r*2\advance\dn*\r* \advance\N*-2\k**#2\relax
\ifnum\l**<6\def\xscale{#4}\def\yscale{#5}\relax
{\angle**0\rotate(#2)}\gmov*(#1,0){\sm*}\advance\k**\dn*
{\angle**0\rotate(\k**)}\gmov*(#1,0){\sl*}\advance\k**\m**
{\angle**0\rotate(\k**)}\gmov*(#1,0){\sm*}\loop
\advance\k**\l**{\angle**0\rotate(\k**)}\gmov*(#1,0){\sl*}\advance\k**\m**
{\angle**0\rotate(\k**)}\gmov*(#1,0){\sm*}\advance\N*-1\ifnum\N*>0\repeat
{\angle**0\rotate(#3)}\gmov*(#1,0){\sl*}\def\xscale{1}\def\yscale{1}\else
\advance\k**\dn* \Earc(#1)[#2,\k**][#4,#5]\loop\advance\k**\m** \r*\k**
\advance\k**\l** {\Earc(#1)[\r*,\k**][#4,#5]}\relax
\advance\N*-1\ifnum\N*>0\repeat
\advance\k**\m**\Earc(#1)[\k**,#3][#4,#5]\fi}}

\def\triangEarc#1(#2)[#3,#4][#5,#6]{{\k**=#3\n*=#4\advance\n*-\k**
\L*=1000sp\L*#2\L* \multiply\L*\n* \multiply\L*\Nhalfperiods
\divide\L*57\N*\L* \divide\N*1000\ifnum\N*=0\N*1\fi
\d**=#2\Lengthunit \d*\d** \divide\d*57\multiply\d*\n*
\r*\n*  \divide\r*\N* \ifnum\r*<2\r*2\fi
\m**\r* \divide\m**2 \l**\r* \advance\l**-\m** \N*\n* \divide\N*\r*
\dt*\d* \divide\dt*\N* \dt*.5\dt* \dt*#1\dt*
\divide\dt*1000\multiply\dt*\magnitude
\k**\r* \multiply\k**\N* \dn*\n* \advance\dn*-\k** \divide\dn*2\relax
\r*\l** \divide\r*2\advance\dn*\r* \advance\N*-1\k**#3\relax
\def\xscale{#5}\def\yscale{#6}\relax
{\angle**0\rotate(#3)}\gmov*(#2,0){\sm*}\advance\k**\dn*
{\angle**0\rotate(\k**)}\gmov*(#2,0){\sl*}\advance\k**-\m**
\advance\l**\m**\loop\dt*-\dt* \d*\d** \advance\d*\dt*
\advance\k**\l**{\angle**0\rotate(\k**)}\rgmov*(\d*,0pt){\sl*}\relax
\advance\N*-1\ifnum\N*>0\repeat\advance\k**\m**
{\angle**0\rotate(\k**)}\gmov*(#2,0){\sl*}\relax
{\angle**0\rotate(#4)}\gmov*(#2,0){\sl*}\def\xscale{1}\def\yscale{1}}}

\def\waveEarc#1(#2)[#3,#4][#5,#6]{{\k**=#3\n*=#4\advance\n*-\k**
\L*=4000sp\L*#2\L* \multiply\L*\n* \multiply\L*\Nhalfperiods
\divide\L*57\N*\L* \divide\N*1000\ifnum\N*=0\N*1\fi
\d**=#2\Lengthunit \d*\d** \divide\d*57\multiply\d*\n*
\r*\n*  \divide\r*\N* \ifnum\r*=0\r*1\fi
\m**\r* \divide\m**2 \l**\r* \advance\l**-\m** \N*\n* \divide\N*\r*
\dt*\d* \divide\dt*\N* \dt*.7\dt* \dt*#1\dt*
\divide\dt*1000\multiply\dt*\magnitude
\k**\r* \multiply\k**\N* \dn*\n* \advance\dn*-\k** \divide\dn*2\relax
\divide\N*4\advance\N*-1\k**#3\def\xscale{#5}\def\yscale{#6}\relax
{\angle**0\rotate(#3)}\gmov*(#2,0){\sm*}\advance\k**\dn*
{\angle**0\rotate(\k**)}\gmov*(#2,0){\sl*}\advance\k**-\m**
\advance\l**\m**\loop\dt*-\dt*
\d*\d** \advance\d*\dt* \dd*\d** \advance\dd*1.3\dt*
\advance\k**\r*{\angle**0\rotate(\k**)}\rgmov*(\d*,0pt){\sl*}\relax
\advance\k**\r*{\angle**0\rotate(\k**)}\rgmov*(\dd*,0pt){\sl*}\relax
\advance\k**\r*{\angle**0\rotate(\k**)}\rgmov*(\d*,0pt){\sl*}\relax
\advance\k**\r*
\advance\N*-1\ifnum\N*>0\repeat\advance\k**\m**
{\angle**0\rotate(\k**)}\gmov*(#2,0){\sl*}\relax
{\angle**0\rotate(#4)}\gmov*(#2,0){\sl*}\def\xscale{1}\def\yscale{1}}}

\newcount\CatcodeOfAtSign
\CatcodeOfAtSign=\the\catcode`\@
\catcode`\@=11
\def\@arc#1[#2][#3]{\rlap{\Lengthunit=#1\Lengthunit
\sm*\l*arc(#2.1914,#3.0381)[#2][#3]\relax
\mov(#2.1914,#3.0381){\l*arc(#2.1622,#3.1084)[#2][#3]}\relax
\mov(#2.3536,#3.1465){\l*arc(#2.1084,#3.1622)[#2][#3]}\relax
\mov(#2.4619,#3.3086){\l*arc(#2.0381,#3.1914)[#2][#3]}}}

\def\dash@arc#1[#2][#3]{\rlap{\Lengthunit=#1\Lengthunit
\d*arc(#2.1914,#3.0381)[#2][#3]\relax
\mov(#2.1914,#3.0381){\d*arc(#2.1622,#3.1084)[#2][#3]}\relax
\mov(#2.3536,#3.1465){\d*arc(#2.1084,#3.1622)[#2][#3]}\relax
\mov(#2.4619,#3.3086){\d*arc(#2.0381,#3.1914)[#2][#3]}}}

\def\wave@arc#1[#2][#3]{\rlap{\Lengthunit=#1\Lengthunit
\w*lin(#2.1914,#3.0381)\relax
\mov(#2.1914,#3.0381){\w*lin(#2.1622,#3.1084)}\relax
\mov(#2.3536,#3.1465){\w*lin(#2.1084,#3.1622)}\relax
\mov(#2.4619,#3.3086){\w*lin(#2.0381,#3.1914)}}}

\def\bezier#1(#2,#3)(#4,#5)(#6,#7){\N*#1\l*\N* \advance\l*\*one
\d* #4\Lengthunit \advance\d* -#2\Lengthunit \multiply\d* \*two
\b* #6\Lengthunit \advance\b* -#2\Lengthunit
\advance\b*-\d* \divide\b*\N*
\d** #5\Lengthunit \advance\d** -#3\Lengthunit \multiply\d** \*two
\b** #7\Lengthunit \advance\b** -#3\Lengthunit
\advance\b** -\d** \divide\b**\N*
\mov(#2,#3){\sm*{\loop\ifnum\m*<\l*
\a*\m*\b* \advance\a*\d* \divide\a*\N* \multiply\a*\m*
\a**\m*\b** \advance\a**\d** \divide\a**\N* \multiply\a**\m*
\rmov*(\a*,\a**){\unhcopy\spl*}\advance\m*\*one\repeat}}}

\catcode`\*=12

\newcount\n@ast
\def\n@ast@#1{\n@ast0\relax\get@ast@#1\end}
\def\get@ast@#1{\ifx#1\end\let\next\relax\else
\ifx#1*\advance\n@ast1\fi\let\next\get@ast@\fi\next}

\newif\if@up \newif\if@dwn
\def\up@down@#1{\@upfalse\@dwnfalse
\if#1u\@uptrue\fi\if#1U\@uptrue\fi\if#1+\@uptrue\fi
\if#1d\@dwntrue\fi\if#1D\@dwntrue\fi\if#1-\@dwntrue\fi}

\def\halfcirc#1(#2)[#3]{{\Lengthunit=#2\Lengthunit\up@down@{#3}\relax
\if@up\mov(0,.5){\@arc[-][-]\@arc[+][-]}\fi
\if@dwn\mov(0,-.5){\@arc[-][+]\@arc[+][+]}\fi
\def\lft{\mov(0,.5){\@arc[-][-]}\mov(0,-.5){\@arc[-][+]}}\relax
\def\rght{\mov(0,.5){\@arc[+][-]}\mov(0,-.5){\@arc[+][+]}}\relax
\if#3l\lft\fi\if#3L\lft\fi\if#3r\rght\fi\if#3R\rght\fi
\n@ast@{#1}\relax
\ifnum\n@ast>0\if@up\shade[+]\fi\if@dwn\shade[-]\fi\fi
\ifnum\n@ast>1\if@up\dshade[+]\fi\if@dwn\dshade[-]\fi\fi}}

\def\halfdashcirc(#1)[#2]{{\Lengthunit=#1\Lengthunit\up@down@{#2}\relax
\if@up\mov(0,.5){\dash@arc[-][-]\dash@arc[+][-]}\fi
\if@dwn\mov(0,-.5){\dash@arc[-][+]\dash@arc[+][+]}\fi
\def\lft{\mov(0,.5){\dash@arc[-][-]}\mov(0,-.5){\dash@arc[-][+]}}\relax
\def\rght{\mov(0,.5){\dash@arc[+][-]}\mov(0,-.5){\dash@arc[+][+]}}\relax
\if#2l\lft\fi\if#2L\lft\fi\if#2r\rght\fi\if#2R\rght\fi}}

\def\halfwavecirc(#1)[#2]{{\Lengthunit=#1\Lengthunit\up@down@{#2}\relax
\if@up\mov(0,.5){\wave@arc[-][-]\wave@arc[+][-]}\fi
\if@dwn\mov(0,-.5){\wave@arc[-][+]\wave@arc[+][+]}\fi
\def\lft{\mov(0,.5){\wave@arc[-][-]}\mov(0,-.5){\wave@arc[-][+]}}\relax
\def\rght{\mov(0,.5){\wave@arc[+][-]}\mov(0,-.5){\wave@arc[+][+]}}\relax
\if#2l\lft\fi\if#2L\lft\fi\if#2r\rght\fi\if#2R\rght\fi}}

\catcode`\*=11

\def\Circle#1(#2){\halfcirc#1(#2)[u]\halfcirc#1(#2)[d]\n@ast@{#1}\relax
\ifnum\n@ast>0\L*=\xscale\Lengthunit
\ifnum\angle**=0\clap{\vrule width#2\L* height.1pt}\else
\L*=#2\L*\L*=.5\L*\special{em:linewidth .001pt}\relax
\rmov*(-\L*,0pt){\sm*}\rmov*(\L*,0pt){\sl*}\relax
\special{em:linewidth \the\linwid*}\fi\fi}

\catcode`\*=12

\def\wavecirc(#1){\halfwavecirc(#1)[u]\halfwavecirc(#1)[d]}

\def\dashcirc(#1){\halfdashcirc(#1)[u]\halfdashcirc(#1)[d]}

\def\xscale{1}
\def\yscale{1}

\def\Ellipse#1(#2)[#3,#4]{\def\xscale{#3}\def\yscale{#4}\relax
\Circle#1(#2)\def\xscale{1}\def\yscale{1}}

\def\dashEllipse(#1)[#2,#3]{\def\xscale{#2}\def\yscale{#3}\relax
\dashcirc(#1)\def\xscale{1}\def\yscale{1}}

\def\waveEllipse(#1)[#2,#3]{\def\xscale{#2}\def\yscale{#3}\relax
\wavecirc(#1)\def\xscale{1}\def\yscale{1}}

\def\halfEllipse#1(#2)[#3][#4,#5]{\def\xscale{#4}\def\yscale{#5}\relax
\halfcirc#1(#2)[#3]\def\xscale{1}\def\yscale{1}}

\def\halfdashEllipse(#1)[#2][#3,#4]{\def\xscale{#3}\def\yscale{#4}\relax
\halfdashcirc(#1)[#2]\def\xscale{1}\def\yscale{1}}

\def\halfwaveEllipse(#1)[#2][#3,#4]{\def\xscale{#3}\def\yscale{#4}\relax
\halfwavecirc(#1)[#2]\def\xscale{1}\def\yscale{1}}

\catcode`\@=\the\CatcodeOfAtSign

\section{Introduction}

Experimental and theoretical investigation of the Lamb shift and hyperfine
structure of muonic hydrogen (${\rm \mu p}$) and muonic deuterium (${\rm \mu d}$)
can lead to essential progress in determining important fundamental
parameters of the proton and deuteron. A better understanding of the
nuclear structure and polarizability effects in these hydrogenic atoms
can be gained due to such studies \cite{BR,BE,SK,CFM,F}.
Namely the proton structure and polarizability corrections lead to main
theoretical uncertainties on the Lamb shift and in the hyperfine splittings
(HFS) of the hydrogen atom. One of the recent studies of these effects in the HFS
in electronic and muonic hydrogen carried out in \cite{AP}
shows that 70 $\%$ of the difference ${\rm (\Delta E_{HFS}^{QED}-
\Delta E_{HFS}^{exp})}$ = 0.046 MHz (QED contribution ${\rm
\Delta E_{HFS}^{QED}}$ doesn't take into account the proton recoil,
structure and polarizability corrections) in the case
of electronic hydrogen can be explained on the basis of effective
field theory describing the interaction of baryons with photons
and leptons. The measurement of the ${\rm 2P\div 2S}$ Lamb shift in muonic
hydrogen has been carried out during recent years at PSI (Paul Scherrer Institute)
\cite{K}. The goal of the experiment is to measure the Lamb shift in muonic
hydrogen with 30 ppm precision and to obtain the root mean square (rms)
proton charge radius with ${\rm 10^{-3}}$ relative accuracy that is an order
of the magnitude better than in the analysis of the elastic e-p scattering
and the Lamb shift in electronic hydrogen. Another important task is connected with
the study of the hydrogen - deuterium isotope shift for the interval ${\rm
1S\div 2S}$ \cite{GB,H,P}. The experimental value of the H-D isotope shift
\begin{equation}
{\rm \Delta E_{H-D}(1S\div 2S)=670~994~334.64(15)~kHz,~~~\delta=2.2\cdot 10^{-10}}
\end{equation}
was obtained with such high precision that the nucleus structure and
polarizability effects should be taken into account on close theoretical
examination of this quantity. Using the result (1) the numerical value for the
difference of the proton and deuteron charge radii can be evaluated as follows
\cite{EGS}:
\begin{equation}
{\rm r_d^2-r_p^2=3.8213(11.7)~fm^2.}
\end{equation}
So, when the Lamb shift measurement in the system ${\rm \mu p}$ with the
precision 30 ppm will be accomplished the deuteron radius ${\rm r_d}$
can be derived by means of (2) with the accuracy ${\rm 10^{-3}}$.
Another approach to determine more
exact value of the deuteron charge radius is related to the measurement
of the muonic hydrogen - muonic deuterium isotope shift for the interval
${\rm 1S\div 2S}$. In this case the precision of the appropriate experiment
should be comparable with (1). The possibility of such measurement of the
${\rm \mu p}$ - ${\rm \mu d}$ isotope shift for the interval ${\rm 1S\div 2S}$ is considered
to be quite real taking into account the relative error of order ${\rm 10^{-5}}$
at PSI for the Lamb shift measurement in muonic hydrogen. Evidently the
calculation of all possible corrections to the Lamb shift in muonic hydrogen
and deuterium with similar precision must be performed. The total value of the
Lamb shift is determined by the sum of quantum electrodynamical (QED)
contributions (one-loop, two-loop, three-loop corrections, recoil corrections,
radiative recoil corrections) and nuclear corrections. Whereas QED contributions
were obtained presently with the relative accuracy ${\rm 10^{-7}}$ due to numerous
calculations \cite{EGS} the nuclear structure corrections are known less
precisely. The accuracy of the corresponding calculation of the nucleus structure
and polarizability effects depends on the experimentally measured
nuclear densities of the charge, magnetic moment, ... and on the used
theoretical models. Both experiments for the isotope shift (1) and the
hydrogen HFS can play the selecting role among numerous nuclear models.

The ground state hyperfine splitting in deuterium represents another important
quantity where the nuclear structure corrections can be tested
experimentally. The experimental value of the deuterium HFS was obtained
with high accuracy many years ago \cite{WR,EGS}:
\begin{equation}
{\rm \Delta E^{exp}_{HFS}(D)=327~384.352~521~9(17)(3)~ kHz,~~~\delta=5.2\times
10^{-12}}.
\end{equation}
The difference between the experimental (3) and
theoretical values for the deuterium HFS accounting
for the QED corrections is equal to ${\rm\Delta E_{HFS}^{exp}(D)}$ -
${\rm \Delta E_{HFS}^{th}(D)}$ = 45 kHz. The essential
contribution to the theoretical quantity ${\rm \Delta E_{HFS}^{th} (D)}$
is given also by the nuclear structure corrections. The deuteron is the spin
1 particle and its electromagnetic structure is described by three form factors. The aim
of our study consists in the exact consideration of the deuteron structure
corrections of order ${\rm (Z\alpha)^5}$ in the Lamb shift and HFS of
the electronic and muonic deuterium. In distinction to previous investigations
of these problems \cite{LS,S,MPK,FP} we use the explicitly covariant approach
to the description of the electromagnetic deuteron - lepton interaction
to construct the quasipotential of the two-photon
interaction. Moreover, we take into account modern experimental data on the
deuteron electromagnetic form factors: charge monopole, charge quadrupole
and magnetic dipole. We leave aside the polarizability effects mentioned above.

\section{Corrections of order $(Z\alpha)^5$ to the deuterium Lamb shift}

The main nuclear structure dependent contribution of order ${\rm (Z\alpha)^4}$
to the hydrogen Lamb shift is determined by the one-photon interaction. In the
one-photon exchange approximation the amplitude of the scattering process
${\rm e d\rightarrow e d}$ is just the contraction of the electron and
deuteron electromagnetic currents, multiplied by the photon propagator. The
parameterization of the deuteron electromagnetic current takes the form
\cite{BP,BH}:
\begin{equation}
{\rm J^\mu_d(p_2,q_2)=\varepsilon^\ast_\rho(q_2)\Biggl\{\frac{(p_2+q_2)_\mu}
{2m_2}g_{\rho\sigma}F_1(k^2)
-\frac{(p_2+q_2)_\mu}{2m_2}\frac{k_\rho k_\sigma}{2m_2^2}F_2(k^2)-
\Sigma^{\mu\nu}_{\rho\sigma}\frac{k^\nu}{2m_2}F_3(k^2)\Biggr\}\varepsilon_\sigma(p_2),}
\end{equation}
where ${\rm p_2, q_2}$ are four momenta of the deuteron in the initial and final states,
${\rm k=q_2-p_2}$, ${\rm m_2}$ is the deuteron mass. The spin 1 polarization
vectors ${\rm \epsilon_\mu}$ satisfy the following conditions:
\begin{equation}
{\rm \varepsilon^\ast_\mu({\bf k},\lambda)\varepsilon^\mu({\bf k},\lambda')=-
\delta_{\lambda\lambda'},~~~ k_\mu\varepsilon^\mu({\bf k},\lambda)=0,~~~
\sum_\lambda\varepsilon^\ast_\mu({\bf k},\lambda)\varepsilon_\nu({\bf k},\lambda)=
-g_{\mu\nu}+\frac{k_\mu k_\nu}{m_2^2}.}
\end{equation}
The generator of the infinitesimal Lorentz transformations
\begin{equation}
{\rm \Sigma^{\mu\nu}_{\rho\sigma}=g^\mu_\rho g^\nu_\sigma-g^\mu_\sigma g^\nu_\rho.}
\end{equation}
The deuteron electromagnetic form factors ${\rm F_i(k^2)}$ depend
on the square of the photon four momentum. They are related to the deuteron
charge ${\rm F_C}$, magnetic ${\rm F_M}$ and quadrupole ${\rm F_Q}$ form
factors by the expressions:
\begin{displaymath}
{\rm F_C=F_1+\frac{2}{3}\eta\left[F_1+(1+\eta)F_2-F_3\right],}
\end{displaymath}
\begin{equation}
{\rm F_M=F_3,~~~\eta=-\frac{k^2}{4m_2^2},}
\end{equation}
\begin{displaymath}
{\rm F_Q=F_1+(1+\eta)F_2-F_3.}
\end{displaymath}
The lepton electromagnetic current has the form:
\begin{equation}
{\rm J_l^\mu(p_1,q_1)=\bar u(q_1)\left[\frac{(p_1+q_1)^\mu}{2m_1}-(1+\kappa_l)
\sigma^{\mu\nu}\frac{k_\nu}{2m_1}\right]u(p_1),}
\end{equation}
where ${\rm p_1, q_1}$ are the electron (muon) four momenta in the initial and final
states, $\sigma^{\mu\nu}$ = $(\gamma^\mu\gamma^\nu-\gamma^\nu\gamma^\mu)/2$,
$\kappa_l$ is the lepton anomalous magnetic moment, ${\rm m_1}$ is the lepton mass.
To obtain the one-photon
interaction contribution to the Lamb shift, it is necessary to average the
currents (4) and (8) over the electron and deuteron spins. As a result the
contribution to the Lamb shift of order ${\rm (Z\alpha)^4}$ is expressed through
the deuteron charge radius ${\rm r_d}$  as follows:
\begin{equation}
{\rm E^{Ls}=\frac{2\mu^3}{3n^3}(Z\alpha)^4\left[r_d^2+F_M(0)-F_C(0)\right],~~~
r_d^2=\frac{6}{F_C(0)}\frac{dF_C(k^2)}{dk^2}\vert_{k^2=0},}
\end{equation}
where ${\rm \mu=m_1m_2/(m_1+m_2)}$ is the reduced mass.
The numerical value of this expression for the interval ${\rm 1S\div 2S}$
in muonic deuterium for the ${\rm r_d=2.094}$ fm is equal to -186.74 meV.
The magnetic quadrupole term of Eq. (9) which is proportional to the
difference ${\rm [F_M(0)-F_C(0)]}$ = ${\rm 2\mu_d-1}$, coincides with the
result of Ref. \cite{KMS}
for the spin 1 particle. Its contribution to the energy spectrum amounts to
0.2 $\%$. Consider the two-photon exchange amplitudes shown in fig.1. They
give the corrections of order ${\rm (Z\alpha)^5}$ to the deuterium Lamb shift.
The spin 1 particle (deuteron) exists in the intermediate state
of the processes of the virtual Compton scattering. The amplitudes of the virtual
Compton scattering on the lepton and deuteron can be presented as follows:
\begin{equation}
{\rm M_{\mu\nu}^{(l)}=\bar u(q_1)\left[\gamma_\mu\frac{\hat p_1+\hat k+m_1}
{(p_1+k)^2-m_1^2}\gamma_\nu+\gamma_\nu\frac{\hat p_1-\hat k+m_1}
{(p_1-k)^2-m_1^2}\gamma_\mu\right]u(p_1)},
\end{equation}
\begin{equation}
{\rm M_{\mu\nu}^{(d)}=\varepsilon^\ast_\rho(q_2)\left[\frac{(q_2+p_2-k)_\mu}
{2m_2}g_{\rho\lambda}F_1-\frac{(q_2+p_2-k)_\mu}{2m_2}\frac{k_\rho k_\lambda}
{2m_2^2}F_2-\Sigma^{\mu\alpha}_{\rho\lambda}\frac{k_\alpha}{2m_2}F_3\right]\times}
\end{equation}
\begin{displaymath}
{\rm \times \frac{-g_{\lambda\omega}+\frac{(p_2-k)_\lambda(p_2-k)_\omega}{m_2^2}}
{(p_2-k)^2-m_2^2}\Biggl[\frac{(p_2+q_2-k)_\nu}{2m_2}g_{\omega\sigma}F_1-
\frac{(p_2+q_2-k)_\nu}{2m_2}\frac{k_\omega
k_\sigma}{2m_2^2}F_2+
\Sigma^{\nu\beta}_{\omega\sigma}\frac{k_\beta}{2m_2}F_3\biggr]\varepsilon_\sigma(p_2)}.
\end{displaymath}

\begin{figure}
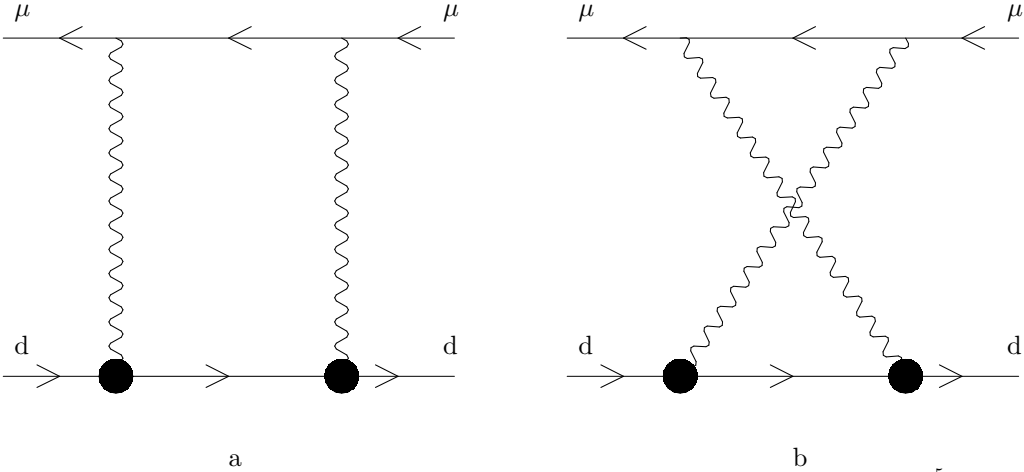

\magnitude=2000
\GRAPH(hsize=15){
\mov(0,0){\lin(1,0)}%
\mov(3,0){\lin(1,0)}%
\mov(0,3){\lin(4,0)}%
\mov(5,0){\lin(1,0)}%
\mov(8,0){\lin(1,0)}%
\mov(5,3){\lin(4,0)}%
\mov(3.,0.){\Circle**(0.3)}%
\mov(6.,0){\Circle**(0.3)}%
\mov(1.,0.){\Circle**(0.3)}%
\mov(8.,0.){\Circle**(0.3)}%
\mov(2.,-0.8){a}%
\mov(7.,-0.8){b}
\mov(0.5,3.){\lin(0.2,0.1)}%
\mov(0.5,3.){\lin(0.2,-0.1)}%
\mov(2.,3.){\lin(0.2,0.1)}%
\mov(2.,3.){\lin(0.2,-0.1)}%
\mov(3.5,3.){\lin(0.2,0.1)}%
\mov(3.5,3.){\lin(0.2,-0.1)}
\mov(5.5,3.){\lin(0.2,0.1)}%
\mov(5.5,3.){\lin(0.2,-0.1)}%
\mov(7.,3.){\lin(0.2,0.1)}%
\mov(7.,3.){\lin(0.2,-0.1)}%
\mov(8.5,3.){\lin(0.2,0.1)}%
\mov(8.5,3.){\lin(0.2,-0.1)}
\mov(0.5,0.){\lin(-0.2,0.1)}%
\mov(0.5,0.){\lin(-0.2,-0.1)}%
\mov(2.,0.){\lin(-0.2,0.1)}%
\mov(2.,0.){\lin(-0.2,-0.1)}%
\mov(3.5,0.){\lin(-0.2,0.1)}%
\mov(3.5,0.){\lin(-0.2,-0.1)}
\mov(5.5,0.){\lin(-0.2,0.1)}%
\mov(5.5,0.){\lin(-0.2,-0.1)}%
\mov(7.,0.){\lin(-0.2,0.1)}%
\mov(7.,0.){\lin(-0.2,-0.1)}%
\mov(8.5,0.){\lin(-0.2,0.1)}%
\mov(8.5,0.){\lin(-0.2,-0.1)}
\mov(1.,0){\lin(2.,0)}%
\mov(6.,0){\lin(2.,0)}%
\mov(1,0){\wavelin(0,3)}%
\mov(3,0){\wavelin(0,3)}
\mov(6,0){\wavelin(2,3)}
\mov(8,0){\wavelin(-2,3)}%
\mov(0.1,0.2){d}%
\mov(0.1,3.2){$\mu$}%
\mov(3.9,0.2){d}%
\mov(3.9,3.2){$\mu$}%
\mov(5.1,0.2){d}%
\mov(5.1,3.2){$\mu$}%
\mov(8.9,0.2){d}%
\mov(8.9,3.2){$\mu$}%
}
\caption{Nuclear structure corrections of order ${\rm (Z\alpha)^5}$.}
\end{figure}

To extract the contribution of the two-photon amplitudes to the Lamb shift
we can average the tensors (10) and (11) over the lepton and deuteron spins.
Multiplying obtained relations and contracting over the Lorentz indices
we can write necessary quasipotential in the form (we
used the system FORM \cite{V} for the trace calculations of the Dirac $\gamma$
matrices and the contraction over the Lorentz indices):
\begin{equation}
{\rm
V_{2\gamma}^{Ls}(ed)=\frac{2m_1(Z\alpha)^2}{3m_2^2}\int\frac{id^4k}{\pi^2}
\frac{1}{(k^2)^2(k^4-4k_0^2m_1^2)(k^2-4k_0^2m_2^2)}\Biggl\{4F_1^2m_2^2k^2(k^2-
k_0^2)\times}
\end{equation}
\begin{displaymath}
{\rm \times\left[3m_2^2-k^2+k_0^2\right]
+4F_1F_2(k^2-k_0^2)^2[k^4-m_2^2(k^2+2k_0^2)]+4F_1F_3m_2^2k^2(k^2-k_0^2)
(k^2-2k_0^2)-}
\end{displaymath}
\begin{displaymath}
{\rm -F_2^2\frac{k^2(k^2-k_0^2)}{m_2^2}\left[k^4-m_2^2(k^2+3k_0^2)\right]
-2F_1F_3k^4(k^2-k_0^2)^2+F_3^2k^2k_0^2[-3k^4+4m_2^2(2k^2+k_0^2)]\Biggr\}.}
\end{displaymath}

After rotating the ${\rm k_0}$ contour of the Feynman loop integration
we integrate Eq. (12) over the four-dimensional Euclidean space using
the relation:
\begin{equation}
{\rm \int d^4k=4\pi\int_0^\infty k^3 dk\int_0^\pi\sin^2\phi \cdot d\phi,~~~
k_0=k\cos\phi}.
\end{equation}
Then the integration over angle $\phi$ in Eq.(12) can be done analytically.
Calculating the matrix element of obtained potential between Coulomb wave
functions we find the following one-dimensional integral representation
for the contribution to the energy spectrum:
\begin{equation}
{\rm E_{2\gamma}^{Ls}=-\frac{\mu^3(Z\alpha)^5}{6m_1^3m_2^5(m_1^2-m_2^2)
\pi n^3}\int_0^\infty\frac{dk}{k}\Biggl\{F_1^2\Biggl[\frac{12m_1^2m_2^4}{k^3}
\left(m_2^2h_1^3-m_1^2h_2^3\right)+\frac{1}{k}\left(m_2^4h_1^5-m_1^4h_2^5\right)+}
\end{equation}
\begin{displaymath}
{\rm
+\left(m_1^2-m_2^2\right)\left(10k^2m_1^2m_2^2+12m_1^2m_2^4+k^4(m_1^2+
m_2^2)\right)\Biggr]+\frac{F_1F_2}{2m_1^2m_2^2}\Biggl[\frac{1}{k}m_2^4h_1^5
\Bigl(2m_1^2m_2^2+}
\end{displaymath}
\begin{displaymath}
{\rm +k^2(2m_1^2-m_2^2)\Bigr)
-\frac{1}{k}m_1^6h_2^5(k^2+2m_2^2)+k^2(m_1^2-m_2^2)\Bigl(-10m_1^4m_2^2+
k^4(m_1^4+m_1^2m_2^2-m_2^4)+}
\end{displaymath}
\begin{displaymath}
{\rm +4k^2(3m_1^4m_2^2-2m_1^2m_2^4)\Bigr)\Biggr]
+\frac{F_2^2}{16m_1^2m_2^4}\Biggl[-h_2^7km_1^6+h_1^5km_2^4(4m_1^2m_2^2+
k^2(4m_1^2-3m_2^2))+}
\end{displaymath}
\begin{displaymath}
{\rm +k^4(m_1^2-m_2^2)\Biggl(-50m_1^4m_2^4+k^4(m_1^4+m_1^2m_2^2-3m_2^4)
+2k^2(7m_1^4m_2^2-13m_1^2m_2^4)\Biggr)\Biggr]+}
\end{displaymath}
\begin{displaymath}
{\rm
+F_1F_3\Biggl[\frac{4m_1^2m_2^2}{k}\left(m_1^2h_2^3-m_2^2h_1^3\right)-
2k\left(m_2^4h_1^3-m_1^4h_2^3\right)-2k^2(m_1^2-m_2^2)\left(8m_1^2m_2^2+k^2(m_1^2+m_2^2)
\right)\Biggr]+}
\end{displaymath}
\begin{displaymath}
{\rm +\frac{F_2F_3}{2m_2^2}\Bigl[h_2^5km_1^4-h_1^5km_2^4-k^4(10m_1^2m_2^2
+k^2(m_1^2+m_2^2))(m_1^2-m_2^2)\Bigr]+}
\end{displaymath}
\begin{displaymath}
{\rm +F_3^2\Bigl[-2h_2^3km_1^4+h_1km_2^2(8m_1^2m_2^2+k^2(3m_1^2-m_2^2))
+k^2(m_1^2-m_2^2)(6m_1^2m_2^2+k^2(2m_1^2-m_2^2))\Bigr]\Biggr\},}
\end{displaymath}
where ${\rm h_i=\sqrt{k^2+4m_i^2}}$. The analysis of the coefficients
with the deuteron form factors in Eq. (14) at small values of the variable
k shows that there are three terms proportional to ${\rm F_1^2}$,
${\rm F_1F_2}$, ${\rm F_1F_3}$ which contain infrared divergences. For the
regularization
of these terms it is necessary to supplement the quasipotential of Eq. (14) by the
iteration term which gives the following contribution to the energy levels:
\begin{equation}
{\rm \Delta E_{iter}^{Ls}=<[V_{1\gamma}\times G^f\times V_{1\gamma}]^{Ls}>=}
\end{equation}
\begin{displaymath}
{\rm =-\frac{16\mu^4(Z\alpha)^5}{\pi n^3}\int_0^\infty\frac{dk}{k^4}
\left[F_1^2(0)+2F_1(0)F'_1(0)k^2+\frac{k^2}{3m_2^2}F_1(0)F_2(0)\right],}
\end{displaymath}
where ${\rm G^{f~-1}=(b^2-{\bf p}^2)/2\mu_R}$ is the inverse free two-particle
propagator \cite{MF}, ${\rm V_{1\gamma}}$ is the quasipotential of the one-photon interaction.
Furthermore to eliminate the divergence in the part ${\rm \sim F_1(k^2)
F_3(k^2)}$ we subtract from it the corresponding contribution accounting for
the point-like deuteron ${\rm \sim F_1(0)F_3(0)}$ (needless to say that the one-loop
corrections of order ${\rm (Z\alpha)^5}$ for the point deuteron in the Lamb
shift must be studied independently \cite{SY}, the relating studies are in
progress). To perform the numerical calculations on the
basis of Eq. (14) we employ recent experimental data on the deuteron
electromagnetic form factors \cite{A} extracted from the tensor polarization
data in the elastic electron - deuteron scattering at given values of the four-
momentum transfer. Some useful parameterizations exist for the deuteron form
factors in the range ${\rm 0\leq k\leq 1.4}$ GeV. We used the parameterization
from Ref. \cite{KS} of the following form:
\begin{equation}
\left(\begin{array}{c}
{\rm F_C} \\
{\rm F_Q} \\
{\rm F_M} \\
\end{array}\right)=F_D^2\left(\frac{k^2}{4}\right)\cdot {\it M(\eta)}\cdot
\left(\begin{array}{c}
{\rm f_0} \\
{\rm f_1} \\
{\rm f_2}  \\
\end{array}\right),
\end{equation}
where ${\rm F_D}$ is the nucleon dipole form factor, ${\it M(\eta)}$
is the matrix of the coefficients, ${\rm\eta=k^2/4m_2^2}$ and
\begin{equation}
{\rm f_m=k^m\sum_{i=1}^4\frac{a_{mi}}{\alpha_{mi}^2+k^2}}.
\end{equation}

The parameters ${\rm a_{mi}}$, ${\rm \alpha_{mi}}$ describing each
electromagnetic form factor can be found in Ref. \cite{KS} (see also
http://www-dapnia.cea.fr/Sphn/T20/Parametrisations). It is necessary to point
out that the parameterization (16) leads to the following value of the
deuteron charge radius ${\rm r_d= 2.094}$ fm \cite{A} which is 2.6 $\%$
smaller than the value ${\rm r_d=2.148}$ fm, obtained on the basis of Eq. (2)
by means of the ${\rm r_p}$ from Ref. \cite{EGS}. The results
of the numerical integration in Eq. (14) are presented in Table 1. The momenta
of the integration which give the main contribution in Eq. (14) have
a characteristic nuclear scale. The obtained correction of order ${\rm (Z\alpha)^5}$
0.53 kHz for the ${\rm 2S\div 1S}$ interval in electronic hydrogen is in
agreement with the result 0.49 kHz of Ref. \cite{FP}. In the case of the muonic
deuterium analogous contribution is equal to 2.57 meV. Then the value of the
nuclear structure corrections to the isotope shift in muonic (electronic)
hydrogen accounting the results of Ref. \cite{KP,MF1} is as follows:
\begin{equation}
{\rm \Delta E^{IS}_{str}(1S\div 2S) }=\Biggl\{{{\rm \mu p-\mu d}~~~1.41~meV
\atop {\rm e p- e d}~~~0.497~kHz}
\end{equation}
It must be taken into account when the comparison with the experimental data
will become available.

\section{Corrections of order $(Z\alpha)^5$ to the deuterium hyperfine
splitting}

To construct the HFS part of the one-loop quasipotential
one needs to keep not only the terms of the operator (6) with ${\rm
\Sigma_{ij}= 2i\varepsilon_{ijk}S_2^k}$ which are proportional to the
deuteron spin ${\rm {\bf S}_2}$. The terms containing the operator ${\rm
\Sigma_{0n}}$, which are expressed through the generator of the Lorentz boosts
${{\bf \Sigma}}$ for the spin 1 particle, also must be taken into
account. There are two possibilities to determine the HFS part
of the interaction operator:\\
1. To keep consistently all the terms, which can contain the spin - spin
interaction operator ${\rm ({\bf S}_1{\bf S}_2)}$ when calculating the
contraction of the amplitudes (10) and (11) over the Lorentz indices.\\
2. To use the special projection operators on the states of the lepton and deuteron
with the total spins 3/2 and 1/2.\\
In this study we carry out covariant construction of the HFS quasipotential
introducing the projection operators ${\rm \hat \pi_{\mu, 3/2}}$ and
${\rm \hat \pi_{\mu, 1/2}}$ for the particles in the initial and final states of the
following form:

\begin{table}
\caption{Nuclear structure corrections of order ${\rm (Z\alpha)^4}$,
${\rm (Z\alpha)^5}$ to the Lamb shift in electronic and muonic deuterium.}
\bigskip
\begin{tabular}{|c|c|c|c|c|c|c|}     \hline
${\rm E^{Ls}}$ & ${\rm 1S~~ (Z\alpha)^4}$ & ${\rm 1S~~ (Z\alpha)^5}$ &
${\rm 2S~~ (Z\alpha)^4}$ & ${\rm 2S~~ (Z\alpha)^5}$ & ${\rm 2S\div 1S~~
(Z\alpha)^4}$ & ${\rm 2S\div 1S~~ (Z\alpha)^5}$  \\  \hline
${\rm ed}~(kHz)$ &${\rm 6.875\cdot 10^3}$ &-0.603 & ${\rm 0.859\cdot 10^3}$ &
-0.075&${\rm -6.016\cdot 10^3}$&0.527 \\  \hline
${\rm \mu d}~(meV)$ &213.42&-2.94 &26.68 &-0.37 &-186.74&2.57  \\  \hline
\end{tabular}
\end{table}

\begin{equation}
{\rm \hat \pi_{\mu, 3/2}=\left[u(p_1)\varepsilon_\mu(p_2)\right]_{3/2}=\Psi_\mu(P)},
\end{equation}
\begin{equation}
{\rm \sum_\lambda\Psi^\lambda_\mu\bar\Psi^\lambda_\nu=\frac{(\hat
P+M)}{2M}
\left(g_{\mu\nu}-\frac{1}{3}\gamma_\mu\gamma_\nu-\frac{2 P_\mu
P_\nu}{3M^2}+ \frac{P_\mu\gamma_\nu-P_\nu\gamma_\mu}{3M}\right),}
\end{equation}
\begin{equation}
{\rm \hat \pi_{\mu, 1/2}=\left[u(p_1)\varepsilon_\mu(p_2)\right]_{1/2}=\frac{i}{\sqrt{3}}\gamma_5\left
(\gamma_\mu-\frac{P_\mu}{M}\right)\Psi(P)},
\end{equation}
where the spin-vector ${\rm \Psi_\mu(P)}$ and the spinor ${\rm \Psi (P)}$
describe the lepton-deuteron states with the total spin 3/2 and 1/2 respectively,
${\rm M=m_1+m_2}$ is the total mass, ${\rm P=p_1+p_2}$.

Multiplying amplitudes (10) and (11)
and taking into consideration the relations (19)- (21) we can represent
the necessary HFS quasipotential by the following expression:
\begin{equation}
{\rm V_{2\gamma}^{HFS}=(Z\alpha)^2\int\frac{id^4k}{\pi^2}\frac{1}{(k^2)^2}
\frac{1}{k^4-4k_0^2m_1^2}\frac{1}{k^4-4k_0^2m_2^2}\times}
\end{equation}
\begin{displaymath}
{\rm
\times\left\{4F_1F_3\left[2(k_0^2+k^2)-\frac{k^4}{m_2^2}\right]k^2{\bf
k}^2+ 2F_2F_3\frac{k^4{\bf
k}^2}{m_2^2}\left(\frac{k^4}{m_2^2}-4k_0^2+{\bf k}^2\right)+
2F_3^2k^2{\bf k}^2\left(k_0^2+\frac{k^4}{m_2^2}\right)\right\}.}
\end{displaymath}
The number of the form factor terms was reduced by half in comparison with
Eq. (14) because the spin-dependent terms of the second particle are proportional
to the form factor ${\rm F_3}$. The expression (22) is less singular than the
operator (14).
The sole infrared divergence of Eq. (22) when ${\rm k\to 0}$ is connected with
the term ${\rm\sim F_1F_3 k^2}$. It can be eliminated entirely by
subtracting the iteration part of the quasipotential which can be
obtained using Eqs. (4) and (8):
\begin{equation}
{\rm \Delta V_{iter}^{HFS}=\left[V_{1\gamma}\times G^f\times V_{1\gamma}\right]^{HFS}=
\frac{32\mu(Z\alpha)^2}{3m_1m_2}({\bf S}_1{\bf S}_2)\int_0^\infty\frac{dk}{k^2}F_1F_3}.
\end{equation}
Subtracting (23) from (22) we can make the analytical integration over the angle
variables in the Euclidean momentum space. The averaging of the obtained expression
over the Coulomb wave
functions will lead to the appearance the factor ${\rm |\psi(0)|^2}$. As a result
the contribution of the two-photon amplitudes to the deuterium HFS can be
written as the one-dimensional integral:
\begin{equation}
{\rm E_{2\gamma}^{HFS}=E_D^F\cdot\delta^{HFS}_{str}=E_D^F\frac{2(Z\alpha)}{\pi
n^3}\int_0^\infty\frac{dk}{k^2}
\Biggl\{\frac{F_3}{F_3(0)}\left[4F_1+\frac{k^2}{m_2^2}(2F_1-F_2-F_3)+\frac{k^4}{m_2^4}F_2\right]\times}
\end{equation}
\begin{displaymath}
{\rm\times\left[\frac{m_1^2m_2}{m_1^2-m_2^2}\left(1+\frac{k^2}{4m_1^2}\right)^{\frac{3}{2}}-
\frac{m_1m_2^2}{m_1^2-m_2^2}\left(1+\frac{k^2}{4m_2^2}\right)^{\frac{3}{2}}+\frac{k^3}
{8m_1m_2}\right]-\frac{F_3k^2}{F_3(0)}\left(F_1+\frac{F_3}{4}+\frac{3}{4}F_2\frac{k^2}
{m_2^2}\right)\times}
\end{displaymath}
\begin{displaymath}
{\rm \times
\left[\frac{m_2}{m_1^2-m_2^2}\left(1+\frac{k^2}{4m_1^2}\right)^{3/2}-
\frac{m_1}{m_1^2-m_2^2}\left(1+\frac{k^2}{4m_2^2}\right)^{3/2}+
\frac{k^3(m_1^2+m_2^2)}{8m_1^3m_2^3}+\frac{3k}{4m_1m_2}\right]-4\mu\Biggr\}},
\end{displaymath}
where the Fermi energy of the ground state hyperfine splitting in  deuterium
\begin{equation}
{\rm E^F_D=2\mu_d\alpha^4\frac{\mu^3}{m_em_p}=326~967.678(4)} ~kHz,
\end{equation}
${\rm \mu_d}$= 0.857~438~228~4(94) is the deuteron magnetic moment in the nuclear
magnetons, $\mu$ is the reduced mass of the deuterium atom, ${\rm m_p}$ is
the proton mass. The numerical integration of Eq. (24) was also done by means of the
deuteron form factor parameterization (16). As a result the contribution of Eq. (24)
to the HFS of the electronic deuterium
\begin{equation}
{\rm E^{HFS}_{2\gamma}(ed)=-34.72}~ kHz.
\end{equation}
The calculation of the deuteron structure corrections to the HFS of the
ground state was performed previously in the analytical form in the zero radius
approximation for the deuteron form factors.
In this approximation the wave function of the deuteron D-state was omitted
and the S-state wave function  was written in the asymptotic form ${\rm
Be^{-\beta r}}$. The contribution obtained in Ref. \cite{MPK} is the following:
\begin{equation}
{\rm \Delta E^{HFS}(ed)=-E^F_D\cdot \Biggl[\alpha \frac{m_e}{3\kappa}(1+2\ln 2)
-\frac{3\alpha}{8\pi}\frac{m_e}{m_p}\ln\frac{\kappa}{m_e}\left(\mu_d-2-
\frac{3}{\mu_d}\right)+}
\end{equation}
\begin{displaymath}
{\rm +\frac{3\alpha}{4\pi}\frac{m_e}{m_p}\ln\frac{\kappa}{m_p}
\frac{1}{\mu_d}\left(\mu_p^2-2\mu_p-3+\mu_n^2\right)\Biggr]
=-21.31} ~kHz.
\end{displaymath}
where $\kappa$ = 45.7 MeV is inverse deuteron size.
The essential growth of the nuclear structure correction of order ${\rm
(Z\alpha)^5}$ to the deuterium HFS in our case as compared with
Ref. \cite{MPK} can be explained by removing the restriction of the zero
radius approximation. The value ${\rm E^{HFS}_{2\gamma}}$
obtained here leads to significant growth of the difference between
the theory and experiment which amounts to 45 kHz without consideration
of the deuteron nuclear structure and polarizability contributions.
In the case of muonic deuterium the corresponding contribution
\begin{equation}
{\rm E^{HFS}_{2\gamma}(\mu d)=-0.925} ~meV.
\end{equation}
The theoretical uncertainty of the results (18), (26), (28) and presented in
Table 1 is determined by the errors of the experimental data for the deuteron
electromagnetic form factors which amounts to 5 $\%$ in the most important
interval ${\rm 0\leq k\leq 0.5 }$ Gev. So, the theoretical error of the
results obtained in this work may amount to 10 $\%$.
It may be useful to compare the relative values of the corrections connected with the
nuclear structure in the energy spectra of light and heavy hydrogen. The main
one-loop contribution to the HFS of electronic (muonic) hydrogen is determined
by the following expression (the Zemach correction) \cite{Z,BY}:
\begin{equation}
{\rm \Delta E=E_F\cdot\delta_Z=E_F\frac{2\alpha\mu}{\pi^2}\int \frac{d{\bf
p}}{(p^2+b^2)^2}\Biggl [\frac{G_E(-{\bf p}^2)G_M(-{\bf
p}^2)}{1+\kappa}-1\Biggr] =E_F(-2\mu\alpha)R_p,~b=\alpha\mu,}
\end{equation}
where ${\rm R_p}$ is the Zemach radius. In the coordinate representation
the Zemach correction is determined by the magnetic moment density ${\rm\rho_M(r)}$
and by the charge density ${\rm \rho_E(r)}$. The value ${\rm R_p}$
can be obtained in the numerical form by using the parameterization
for the proton electromagnetic form factors obtained at Mainz 20 years ago
from the analysis of elastic electron - proton scattering \cite{Simon}.
The relative contributions of the Zemach correction to the hydrogen HFS
are as follows:
\begin{equation}
electronic~ hydrogen:~~~{\rm R_p}=1.067~fm,~\delta_Z=-40.3~{\rm ppm},
\end{equation}
\begin{equation}
muonic~ hydrogen:~~~{\rm R_p}=1.064~fm,~\delta_Z= -74.7\cdot 10^{-4}.
\end{equation}
The relative deuteron structure contributions obtained
in this study have the following numerical values:
\begin{equation}
electronic~ deuterium~~~{\rm \delta^{HFS}_{str}=-106.19~ppm}
\end{equation}
\begin{equation}
muonic~ deuterium~~~{\rm \delta^{HFS}_{str}=-188.37\cdot 10^{-4}.}
\end{equation}
Three times growth of the contributions (32), (33) as compared with the
expressions (30), (31)
can be caused by increasing the distribution region of the deuteron electric
charge and magnetic moment.
The numerical results obtained in this work for the muonic
deuterium must be taken into consideration when both extracting the deuteron
charge radius in future experiments on the isotope shift
${\rm (\mu p)- (\mu d)}$, and comparing theoretical predictions with
measurements of the HFS in deuterium.

\begin{acknowledgements}
We are grateful to I.B. Khriplovich and R.A. Sen'kov for useful discussions
and to J. Ball and J. Jourdan for sending us the numerical values
of the parameters for the deuteron electromagnetic form factors.
The work was performed under the financial support of the Program
"Universities of Russia" (grant UR.01.02.016).
\end{acknowledgements}

\end{document}